\documentclass[onecolumn,aps,12pt]{revtex4}
\usepackage{amsmath,graphicx,axodraw}
\def\,{\ifmmode\mskip\thinmuskip\else\leavevmode\thinspace\fi}

\newcommand{\Ree}{\mbox{Re}}
\newcommand{\dd}{\mbox{d}}
\newcommand\ba{\begin{eqnarray}}
\newcommand\ea{\end{eqnarray}}
\newcommand\nn{\nonumber}

\begin{document}

\title{Radiative muon pair production
in high energy electron-positron annihilation process}

\author{A.B.~Arbuzov, V.V.~Bytev, E.A.~Kuraev}
\affiliation{Joint Institute for Nuclear Research, 141980 Dubna,
Russia}

\date{\today}

\begin{abstract}

The lowest order radiative correction to the differential cross-section of 
process of muon pair production with emission of hard photon 
at high energy electron-positron annihilation are calculated.
Taking into account the emission of additional soft and hard photon
the cross-section can be put in the form of Drell-Yan process
in leading logarithmical approximation. Applying the crossing
transformation we obtain the cross section of radiative 
electron-muon high-energy scattering process.
Virtual and soft photon emission contributions of non-leading form are 
tabulated for several typical kinematical
points. The limit of small invariant mass of muon pair is in
agreement with our previous analysis.

\end{abstract}

\maketitle

\section{Introduction}

Process of muon pair production as well as radiative muon pair production 
at high energy in electron-positron collisions is commonly used for calibration
purposes. One of the motivations  of our investigation is the high theoretical
accuracy required for description of differential cross-section.
An additional interest appears in the case of small invariant mass of th muon pair.
For this case the radiative muon pair production is provided by the initial state
hard photon emission kinematics. It can be used as a calibration process
in studying the hadronic systems of small invariant masses created by virtual photon.
The lowest order radiative corrections (RC) in that kinematics to Born cross-section 
as well as the leading  logarithmic (LL)
and next-to-leading (NL) contributions in all orders of perturbation theory (PT)
were considered in our recent paper \cite{smalls1}.

Besides the practical applications, we pursue the another aim in this paper.
The problem is to check the validity 
renormalization group (RG) predictions concerning hard processes of type $2\to 3$.

 Basing on exact (with power accuracy $O(M^2_\mu/s)$)
calculations we confirm the Drell-Yan form of  the cross-section of radiative 
muon pair production in LL. Estimation of non-leading contributions for several 
kinematics points are given as well.

In conclusion we put the cross-section for crossing processes:
radiative electron-muon scattering and muon pair production by photon on electron in LL.

\section{Born cross-section and RC}

In this paper for the process 
\ba
e^+(p_+)+e^-(p_-)\to \mu^+(q_+)+\mu^-(q_-)+\gamma(k_1) 
\ea
we use the following kinematics:
\begin{gather}
\chi_\pm=2k_1p_\pm,\quad\chi_\pm'=2k_1q_\pm,\quad s=(p_-+p_+)^2,\quad
s_1=(q_-+q_+)^2,\nonumber \\
t=(p_--q_-)^2,\quad t_1=(p_+-q_+)^2,\quad u=(p_--q_+)^2,\quad
u_1=(p_+-q_-)^2, \\ \nonumber
p_\pm^2=m^2,\quad  q_\pm=M^2,
\end{gather}
where $M(m)$ is  muon (electron) mass. 
Here all kinematical invariants are much larger than muon  (electron) mass, but 
we take into account terms of order $\ln(M/m)$:
\begin{gather}
s_1\sim s\sim-t\sim-t_1\sim-u\sim-u_1\sim\chi_\pm\gg M^2 \gg m^2, \\ \nonumber
s+s_1+t+t_1+u+u_1=0.
\end{gather}
The differential cross-section of the process with the lowest order radiative
correction (RC) has the  form:
\ba
\frac{\dd \sigma_0}{\dd\Gamma}=\frac{\alpha^3}{2\pi^2s}
m_0\biggl[1+\frac{\alpha}{\pi}(\Delta_{vac}+\Delta_{ff}
+\Delta_{vert}+\Delta_{box}+\Delta_{soft})\biggr], \\ \nonumber
\dd\Gamma=\frac{\dd^3q_+\dd^3q_-\dd^3k_1}{\varepsilon_+\varepsilon_-\omega_1}\delta^4
(p_++p_--q_+-q_--k_1).
\ea
It's convenient to separate starting from Born level definite contributions
from hard photon emission by electron, muon block and their interference:
\ba
m_0=m_0^e+m_0^\mu+m_0^{int}
\ea 
where \cite{berends}
\begin{gather}
m_0^e=A\frac{s}{\chi_-\chi_+}, \quad m_0^\mu=A\frac{s_1}{\chi_-^{'}\chi_+^{'}},\quad
m_0^{int}=A\biggl[-\frac{t}{\chi_-\chi_-^{'}}-\frac{t_1}{\chi_+\chi_+^{'}}
+\frac{u_1}{\chi_+\chi_-^{'}}+\frac{u}{\chi_-\chi_+^{'}}\biggr], \\ \nonumber
A=\frac{t^2+t_1^2+u^2+u_1^2}{ss_1}.
\end{gather}

The standard evaluation of additional soft photon emission contribution gives:
\ba
\frac{\dd\sigma_{soft}}{\dd\sigma_0}=-\frac{\alpha}{4\pi^2}
\int\frac{\dd^3k_2}{\omega_2}\biggl(-\frac{p_-}{p_-k_2}+\frac{p_+}{p_+k_2} 
+\frac{q_-}{q_-k_2}-\frac{q_+}{q_+k_2}\biggr)^2\biggr|_{\omega_2<\Delta\varepsilon\ll \varepsilon}\\ \nonumber
=\frac{\alpha}{\pi}(\Delta_s^e+\Delta_s^\mu+\Delta_s^{int})=\frac{\alpha}{\pi}\Delta_{soft}.
\ea
Here we denote:
\begin{gather}
\Delta_s^e=2(\rho_s+L-1)\ln\frac{m\Delta\varepsilon}{\lambda\varepsilon}+\frac{1}{2}
(\rho_s+L)^2-\frac{\pi^2}{3}, \\ \nonumber
\Delta_s^\mu=2(\rho_{s_1}-L-1)\ln\frac{M\Delta\varepsilon}{\lambda\sqrt{\varepsilon_+\varepsilon_-}}+\frac{1}{2}
(\rho_{s_1}-L)^2-\frac{1}{2}\ln^2\frac{\varepsilon_+}{\varepsilon_-}-\frac{\pi^2}{3}
+\mbox{Li}_2\biggl(\frac{1+c}{2}\biggr), \\ \nonumber
\Delta_s^{int}=\frac{1}{2}(\rho_{t_1}+\rho_u)\ln\frac{t_1}{u}
+\frac{1}{2}(\rho_{t}+\rho_{u_1})\ln\frac{t}{u_1}
+2\ln\frac{t_1}{u}\ln\frac{\sqrt{mM}\Delta\varepsilon}{\lambda\sqrt{\varepsilon\varepsilon_+}}
+2\ln\frac{t}{u_1}\ln\frac{\sqrt{mM}\Delta\varepsilon}{\lambda\sqrt{\varepsilon\varepsilon_-}} \\ \nonumber
+\mbox{Li}_2\biggl(\frac{1+c_-}{2}\biggr)
+\mbox{Li}_2\biggl(\frac{1-c_+}{2}\biggr)
-\mbox{Li}_2\biggl(\frac{1+c_+}{2}\biggr)
-\mbox{Li}_2\biggl(\frac{1-c_-}{2}\biggr),
\end{gather}
where
\begin{gather}
L=\ln\frac{M}{m},\quad \rho_\lambda=\ln\frac{mM}{\lambda^2},\quad \rho_s=\ln\frac{s}{mM} \quad
\rho_{s_1}=\ln\frac{s_1}{mM}, 
\quad \rho_{t}=\ln\frac{-t}{mM},\\ \nonumber \rho_{t_1}=\ln\frac{-t_1}{mM},\quad
\rho_{u}=\ln\frac{-u}{mM},\quad \rho_{u_1}=\ln\frac{-u_1}{mM},\quad
\mbox{Li}_2(z)=-\int\limits_0^z\frac{\dd x}{x}\ln(1-x), \\ \nonumber
c_\pm=\cos(\vec{p}_-\vec{q}_\pm),\quad c=\cos(\vec{q}_+\vec{q}_-).,
\end{gather}
and $\varepsilon, \varepsilon_\pm$ are the energies (in cms) of electron,
muon and $\lambda$ is "photon mass".

Let us now consider RC arising from the Dirac form factor of leptons and  vacuum polarization, 
(the Pauli form factor contribution is suppressed in our kinematics).
They are:
\ba
\Delta_{ff}+\Delta_{vac}=\frac{2m_0^e+m_0^{int}}{m_0}\biggl(\Ree\Gamma(\frac{s_1}{M^2})+\Ree\Pi(s_1)\biggr)
+\frac{2m_0^\mu+m_0^{int}}{m_0}\biggl(\Ree\Gamma(\frac{s}{m^2})+\Ree\Pi(s)\biggr)
\ea
with
\ba
\Ree\Gamma(\frac{s}{m^2})=(\ln\frac{m}{\lambda}-1)(1-\rho_s-L)-\frac{1}{4}(\rho_s+L)^2-\frac{1}{4}(\rho_s+L)
+\frac{\pi^2}{3},\\ \nonumber
\Ree\Gamma(\frac{s_1}{M^2})=(\ln\frac{M}{\lambda}-1)(1-\rho_{s_1}+L)-\frac{1}{4}(\rho_{s_1}-L)^2
-\frac{1}{4}(\rho_{s_1}-L)+\frac{\pi^2}{3}, \\ \nonumber
\Ree\Pi(s_i)=\Ree\Pi^e(s_j)+\Ree\Pi^\mu(s_j)+\Ree\Pi^\tau(s_j)+\Ree\Pi^h(s_j), \\ \nonumber
\Ree\Pi^e(s_j)=\frac{1}{3}(\rho_{s_j}+L)-\frac{5}{9}, \quad
\Ree\Pi^\mu(s_j)=\frac{1}{3}(\rho_{s_j}-L)-\frac{5}{9}.
\ea
Here $s_j$ is the kinematical invariant $s$ or $s_1$.
The contributions from the vacuum polarization from the heavy lepton $\tau$ 
and hadrons $\Pi^\tau, \Pi^h$ are given in \cite{Eidelman:1995ny}.

\section{Calculations of box-type RC}

\begin{figure} 
\includegraphics[scale=1.2]{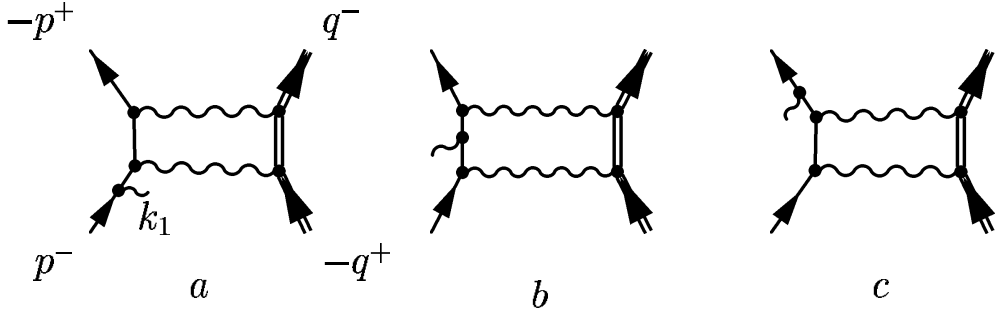}
\caption{Set of box-type FD used in calculation.}
\label{fig:1}
\end{figure}

Consider now amplitudes arising from box-type Feynman Diagrams (FD). There are twelve FD of such a kind,
or $48$ squared matrix elements.
In calculation we restrict ourselves by consideration of only three of box-type FD. Really the total 
contribution of interference of box-type and Born amplitudes can be expressed in form:
\ba
\Ree\Sigma M_{box}M_0^\star=(1+P_1)[(1-P_2) B^e(M_0^e)^*+
(1+P_2) B^e(M_0^\mu)^*],
\ea
with $ M_0^e+M_0^\mu=M_0$, $M_0^e(M_0^\mu)$-are electron (muon) block emission part of the 
Born matrix element; $B^e$-is the electron emission part of contribution to the box-type
amplitude with uncrossed photon legs (see Fig.\ref{fig:1}). Note that calculating the $B^e$ we must 
consider the pentagon type FD (see Fig.\ref{fig:1},b) and two remaining ones (see FD Fig.\ref{fig:1}a,c).

The substitution operators $P_{1,2}$ work as
\ba
P_1 f(p_+,p_-;q_+,q_-,k_1)=f(q_+,q_-;p_+,p_-;-k_1);\nn \\
P_2 f(p_+,p_-;q_+,q_-,k_1)=f(p_+,p_-;q_-,q_+,k_1).
\ea

The operator $P_1$ "changes" the photon emission from electron line to muon line.
The application of operator $P_2$ permits to obtain the contribution from FD at Fig\ref{fig:1}
FD with crossed virtual photon lines.
As a result we obtain:
\ba
\Delta_{box}=-(\rho_s+\rho_\lambda)\ln\frac{tt_1}{uu_1}+\Delta_B^{NL}.
\ea
The expression for $\Delta_B^{NL}$ is rather cumbersome.
The whole contribution to $\Delta_{NL}$ (which does not contains large logarithms) 
would be given in form of the table below.

\section{Vertex-type FD}

\begin{figure} 
\includegraphics[scale=1.2]{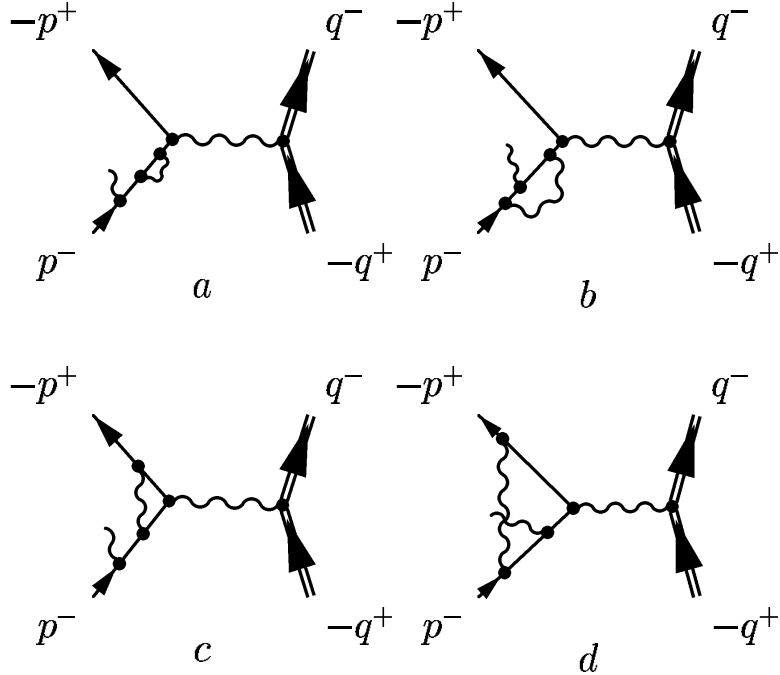}
\caption{Set of vertex FD used in our calculation.}
\label{fig:2}
\end{figure}

Let now consider the contribution arising from FD with vertex type insertions $V^e$
(see Fig.\ref{fig:2}). The other vertex contributions appear from
 this ones by using substitutions.
\ba
\Ree\Sigma M_{vertex}M_0^\star=(1+P_1)(1+P_3) V^e(M_0^e)^*,
\ea
with operator $P_3$ defined as:
\ba
P_3 f(p_+,p_-;q_+,q_-,k_1)=f(p_-,p_+,q_+,q_-;,k_1).
\ea
 The total answer for vertex-type contribution reads:
\ba
\Delta_{vert}=-\frac{1}{2}\frac{m_e+\frac{1}{2}m_i}{m_0}[(\rho_s+L)^2+2(\rho_s+L)(\rho_\lambda+L)-
3(\rho_s+L)+\Delta^{NL}_v(s)] \nn \\
-\frac{1}{2}\frac{m_\mu+\frac{1}{2}m_i}{m_0}[(\rho_{s_1}-L)^2+2(\rho_{s_1}-L)(\rho_\lambda-L)-
3(\rho_{s_1}-L)+\Delta^{NL}_v(s_1)].
\ea

\section{Master-formula}
Extracting  the explicate dependence on vacuum polarization in the form $\frac{1}{|1-\Pi|^2}$
and collecting the leading and non-leading terms arising from soft photon emission,
vertex and box-type FD contributions, as well as lepton form-factors we arrive to
the formula:
%
\begin{gather}
\Delta_{soft}+
\Delta_{box}+\Delta_{vert}+\Delta_{ff}=\Delta_{lead}+\Delta_{NL}.
\end{gather}
 This expression is free from the infrared singularities as 
well as from squares of large logarithms. The form of $\Delta_{lead}$ is consistent 
with renormalization group prescriptions:
\ba
1+\frac{\alpha}{\pi}\Delta_{lead}=(1+\frac{\alpha}{2\pi}\ln\frac{s}{m_e^2}P_\Delta(\varepsilon))^2
(1+\frac{\alpha}{2\pi}\ln\frac{s_1}{M^2}P_\Delta(\varepsilon_+))
(1+\frac{\alpha}{2\pi}\ln\frac{s_1}{M^2}P_\Delta(\varepsilon_-)) \\ \nonumber
+\mbox{O}(\alpha^2),
\ea
with $P_\Delta$ being the $\delta$ part of the kernel of evolution equation:
\ba
P_\Delta(\varepsilon)=2\ln\frac{\Delta\varepsilon}{\varepsilon}+\frac{3}{2}, \nn \\
P_\Delta(\varepsilon_\pm)=2\ln\frac{\Delta\varepsilon}{\varepsilon_\pm}+\frac{3}{2}.
\ea
An additional hard photon emission contribution in leading logarithmical order
can be taken into account using the quasi-real electron's method \cite{Baier}.
It results in the replacement $P_\Delta$ by the whole kernel of evolution 
equation of twist 2 operators 
\ba
P(z)=P^{(1)}(z)=\lim_{\Delta\to 0}[P_\Delta\delta(1-z)+P_\Theta(z)],\nn \\
P_\Delta=2\ln\Delta+\frac{3}{2}, \quad
P_\Theta(z)=\Theta(1-\Delta-z)\frac{1+z^2}{1-z}.
\ea
As a result we arrive to compact form of the cross section:
\ba
\frac{d\sigma^{e^+e^-\to\mu^+\mu^-\gamma}(p_-,p_+,q_-,q_+,k_1)}{d\Gamma}=\int\limits_{x_m}^1\dd x_1
\int\limits_{x_m}^1\dd x_2\int\limits_{y_-}^1\frac{\dd z_-}{z_-}\int\limits_{y_+}^1
\frac{\dd z_+}{z_+}D_e(x_1,s)D_e(x_2,s)\times \\ \nonumber
D_\mu(\frac{y_-}{z_-},s_1)
D_\mu(\frac{y_+}{z_+},s_1)\frac{(1+\frac{\alpha}{\pi}K)}{|1-\Pi(sx_1x_2)|^2}
\frac{\dd \sigma^{e^+e^-\to\mu^+\mu^-\gamma}(x_1p_-,x_2p_+,Q_-,Q_+,k_1)}{d\Gamma_1},
\\ \nonumber
Q_\pm=\frac{z_\pm}{y_\pm}q_\pm, \quad y_\pm=\frac{\varepsilon_\pm}{\varepsilon},\quad\quad\quad\quad\quad
\quad\quad\quad\quad\quad
\ea
and the structure functions $D(x,s)$ having the standard form:
\ba
D_e(x,s)=\delta(1-x)+\frac{\alpha}{2\pi}P^{(1)}(x)\ln\frac{s}{m^2}+...\,\,\, , \nn \\
D_\mu(y,s_1)=\delta(1-y)+\frac{\alpha}{2\pi}P^{(1)}(y)\ln\frac{s_1}{M^2}+... \,\,\, .
\ea
The phase volumes entering the left and right parts of master equation are 
different:
\begin{gather}
\dd\Gamma=\frac{\dd^3q_-}{\varepsilon_-}\frac{\dd^3q_+}{\varepsilon_+}\frac{\dd^3k_1}{\omega_1}\delta(p_++p_--q_+-q_--k_1), \nn \\
\dd\Gamma_1=\frac{\dd^3Q_-}{E_-}\frac{\dd^3Q_+}{E_+}\frac{\dd^3k_1}{\omega_1}\delta(x_2p_++x_1p_--Q_+-Q_--k_1),\\ \nonumber
E_\pm=\frac{z_\pm}{y_\pm}\varepsilon_\pm.
\end{gather}
The lower limits of the energy fractions integrations $x_m,y_m$ are determined by the 
experiment set-up conditions.
The quantity $K$ (so called $K$-factor) collects all the nonleading contributions.
It has contributions from virtual, soft and hard photon emission  terms.
In the Table below we give its value for typical experimental points of the considered process 
 keeping all contributions except ones arising from additional
hard photon emission.

\section{conclusion}

Our consideration was devoted to the lowest order RC. Nevertheless  
result obtained reveals the lowest order expansion of the structure functions $D$.
So the general Drell-Yan form of cross section is established, which is valid in all 
orders of PT. The order of magnitude of nonleading terms can be estimated from the
Table $1$:
\vspace{1cm}
\begin{center}
\begin{tabular}{||c||c|c|c|c|c||}
\hline
\hline
  $N$& $e_-$ &$e_+$& $c_-$& $c_+$ & $\Delta_{NL}$  \\
\hline
\hline
1 & 0.59 & 0.66 & 0.29 & -0.06 &6.77 \\
\hline
2 & 0.67 & 0.67 & 0.50 & 0.30 & 3.24  \\
\hline
3 & 0.68 & 0.65 & 0.69 & -0.50 & 8.68  \\
\hline
4 & 0.59 & 0.56 & -0.30 & -0.30 & 8.35  \\
\hline
\hline
\end{tabular}
\end{center}
Table $1$. Numerical estimation of $\Delta_{NL}$, part of $K$-factor excluding non-leading terms 
arising from hard non-collinear  photon emission (which depends on experimental
set-ups) and the terms proportional to $\ln\Delta\varepsilon/\varepsilon$,
$\ln\Delta\varepsilon/\varepsilon_\pm$ arising from soft photon emission.
\vspace{0.5 cm}

Without  additional calculations we can obtain by the analogy with the result given 
above the cross section
of crossing process - radiative electron-muon scattering:
\ba
e_-(p_1)+\mu_-(q_1)\to e_-(p_2)+\mu_-(q_2)+\gamma(k_1)+(\gamma).
\ea
It can be constructed in complete analogy with the Drell-Yan form of cross section
of above considered process $e_+e_-\to \mu_+\mu_-\gamma$, using in right hand
side as a hard subprocess the Born cross section:
\ba
\frac{d\sigma^{e\mu\gamma}_B}{d\Gamma_{e\mu\gamma}}(p_1,q_1;p_2,q_2,k_1)=\frac{\alpha^3}{16\pi^2(p_1q_1)}
\frac{(p_1q_2)^2+(p_1q_1)^2+(p_2q_1)^2+(p_2q_2)^2}{(p_1p_2)(q_1q_2)}W,
\ea 
with
\ba
d\Gamma_{e\mu\gamma}=\frac{d^3q_2}{q_{20}}\frac{d^3p_2}{p_{20}}\frac{d^3k_1}{\omega_1}\delta^4(p_1+q_1-p_2-q_2-k_1); \nn \\
W=-\biggl(\frac{p_1}{p_1k_1}+\frac{q_1}{q_1k_1}-\frac{p_2}{p_2k_1}-\frac{q_2}{q_2k_1}\biggr)^2.
\ea
It is worth to note that the value of K-factor for the last process is not known.

All the 1-loop integrals used of scalar,vector and tensor types were published in our
previous papers \cite{BKS}.
It's important to note that numerical values of nonleading terms for process
of radiative muon pair production for the case of small muon invariant mass
we find completely in agreement with the result of our paper devoted to this 
kinematical situation \cite{smalls1}, where it was calculated analytically.

\section{Acknowledgement}

We are grateful to grant RFFI 03-02-17077 and two of us E.K. and B.V.to  INTAS 00-00-366.

\end{document}